\documentclass[twocolumn,epjc3]{svjour3}  
\usepackage[T1]{fontenc}
\usepackage[utf8]{inputenc}
\usepackage{lmodern} 
\usepackage[english]{babel}

\usepackage{graphicx,amssymb}
\usepackage{hyperref}

\def \d {{\rm d}}
\def \e {e}
\def \boldk {\mbox{\boldmath$k$}}

\begin{document}

\title{Robinson--Trautman solution with nonlinear electrodynamics}

\author{T. Tahamtan\thanksref{e1,addr1}
        \and
        O. Sv\'{\i}tek\thanksref{e2,addr1}}
\thankstext{e1}{e-mail: tahamtan@utf.mff.cuni.cz}
\thankstext{e2}{e-mail: ota@matfyz.cz}
\institute{Institute of Theoretical Physics, Faculty of Mathematics and Physics, Charles University in Prague, V~Hole\v{s}ovi\v{c}k\'ach 2, 180~00 Prague 8, Czech Republic\label{addr1}}

\date{\today}

\maketitle

\begin{abstract}
Explicit Robinson--Trautman solutions with electromagnetic field satisfying nonlinear field equations are derived and analyzed. The solutions are generated from the spherically symmetric ones. In all studied cases the electromagnetic field singularity is removed while the gravitational one persists. The models resolving curvature singularity in the spherically symmetric spacetimes could not be generalized to Robinson--Trautman geometry using the generating method developed in this paper which indicates that the removal of a singularity in the associated spherically symmetric case might be just a consequence of high symmetry. We show that the obtained solutions are generally of algebraic type II and reduce to type D in spherical symmetry. Asymptotically they tend to the spherically symmetric case as well.
\end{abstract}

\section{Introduction}
Robinson--Trautman spacetimes represent an important class of geometries defined by the requirement of possessing an expanding, nontwisting and nonshearing null geodesic congruence \cite{RobinsonTrautman:1960,RobinsonTrautman:1962,Stephanietal:book,Griffiths:book}. This class contains non-spherical generalizations of black holes without rotation (Schwarzschild solution is a special case in this class, unlike the Kerr black hole). In general, Robinson--Trautman spacetimes do not posses any Killing vectors thus providing solutions devoid of symmetry. Another important aspect of this family is the presence of gravitational radiation connected with the dynamical nature of general solutions within this class. Many global properties of this class in four dimensions have been studied analytically, especially in the last 25 years. In particular, based on the only nontrivial Einstein equation (so-called Robinson--Trautman equation) the asymptotic evolution and global structure of vacuum Robinson--Trautman spacetimes of type~II with spherical topology were investigated by Chru\'{s}ciel and Singleton \cite{Chru1,Chru2,ChruSin}. They showed that characteristic initial value problem for generic, arbitrarily strong smooth initial data converges asymptotically in retarded time to corresponding Schwarzschild metric near its future horizon. Extensions across such ``Schwarzschild-like'' future event horizon are only of a finite order of smoothness. These results were later extended to cover the presence of a cosmological constant which naturally modifies the asymptotic behavior and the solutions tend to Schwarzschild--(anti-)de Sitter \cite{podbic95,podbic97}. Finally, the  Chru\'{s}ciel--Singleton analysis was used for Robinson--Trautman spacetimes admitting additionally pure radiation \cite{BicakPerjes,PodSvi:2005}, where the asymptotic state is described by spherically symmetric \mbox{Vaidya--(anti-)de~Sitter} metric. The location of past quasilocal horizon (which cannot be determined as an event horizon due to the impossibility to extend the solutions to past null infinity) together with the proof of its existence and uniqueness for the vacuum Robinson--Trautman solutions have been studied by Tod \cite{tod}. Later, Chow and Lun \cite{chow-lun} further analyzed properties of this horizon and made numerical study of both the horizon equation and the Robinson--Trautman equation. The analytic results were generalized to nonvanishing cosmological constant \cite{PodSvi:2009}. The relation between asymptotic momentum and local horizon curvature in Robinson--Trautman class was used in the analytic explanation of an "antikick" appearing in numerical studies of asymmetric binary black hole merger \cite{rezzolla}. Recently, solution with minimally coupled free scalar field was derived in \cite{Tahamtan-Svitek-PRD} and shown to posses singularity which is initially naked and only later gets covered by horizon.

\sloppy

Type D vacuum solutions of Robinson-Trautman family contain, apart from Schwarzschild solution, also the C-metric \cite{Krtous} representing uniformly accelerated pair of black holes. C-metric is as well a natural future asymptotics in case of non-smooth initial data for certain subclass of Robinson-Trautman spacetime \cite{Hoenselaers}. By leaving out the usual spherical topology assumption one can obtain a special case of Kasner metric \cite{Stephanietal:book}. Including null radiation source into type D leads to, e.g. Vaidya solution or Kinnersley's rocket \cite{Kinnersley} which is interpreted as an object propelled by emitting directional null radiation (these "rocket" solutions can be generalized to type II). All type D null radiation metrics are known \cite{Frolov}. 

\fussy

Vacuum type N solutions which correspond to spacetimes containing just gravitational radiation have singularity at each wave surface which combine into singular lines \cite{Stephanietal:book,Griffiths:book}. The general solution was given by Foster and Newman \cite{Foster} and they are frequently used to form various sandwich-type waves \cite{Griffiths}. There are no pure radiation solutions of type N. 

There is also a higher-dimensional generalization of Robinson--Trautman spacetimes (containing aligned pure radiation and a cosmological constant) which, however, lacks the rich dynamics present in four dimensions \cite{podolsky-ortaggio}. The existence of horizons (which have generally richer topology than in four dimensions) in this case was subsequently analyzed in \cite{Svitek2011}. These higher-dimensional solutions were as well generalized to admit a source in the form of p-form fields \cite{ortaggio1}. 

Robinson--Trautman spacetimes with Maxwell field were already derived in the founding paper of this family of solutions \cite{RobinsonTrautman:1962}. Later they were studied more extensively \cite{Lind,Newman,Kozameh1}. Among the special cases of these solutions belong Reissner--Nordstr\"{o}m black hole and charged C-metric but it was also shown that this subfamily suffers from non well-posedness \cite{Kozameh2}. The higher-dimensional generalization of Robinson--Trautman spacetimes coupled to Maxwell electrodynamics was derived in \cite{ortaggio2}.

Nonlinear Electrodynamics (NE) was founded almost a century ago and used mainly as a solution to the problem of divergent field of a point charge in the vicinity of its position (see e.g. \cite{Dirac}) also giving satisfactory self-energy of charged particle. The best-known and frequently used form of the theory was introduced already in 1934 by Born and Infeld \cite{BornInfeld}. Nice overview with a lot of useful information was given in a book by Pleba\'{n}ski \cite{Plebanski}. More recently, the attention to Nonlinear Electrodynamics was increased thanks to the discovery that string-generated corrections to Maxwell field have the form of original Born--Infeld theory \cite{Wiltshire}. However, it was also noted that the electric displacement vector in Born--Infeld model has two possible values for single value of electric field. This non-uniqueness was soon solved by adding the so called Hoffmann term \cite{Hoffmann} in Born--Infeld Lagrangian. Additionally, this new model was also used to resolve the spacetime singularity in spherically symmetric case \cite{TT-PhysLettA}.

Later, other NE models were considered for both solving the point charge singularity and resolving the spacetime singularity \cite{Ayon-Beato1,Ayon-Beato2}. Note, however, that these results were obtained in the Hamiltonian framework and their Legendre transform is far from trivial \cite{Bronnikov}. One important example is the so-called Bardeen black hole \cite{Bardeen} which was originally discovered as a regular solution with horizon generated by certain stress energy tensor. Only later this source was interpreted as being created by a specific model of NE \cite{Ayon-Beato3} which, however, does not have a Maxwell limit (in weak field regime).

\sloppy

Many spherically symmetric solutions of Einstein equations with NE were studied, mostly with the Born--Infeld Lagrangian \cite{BI}, logarithmic Lagrangians \cite{Soleng,Hendi-Rad}, square root lagrangian \cite{square root}, power Maxwell models \cite{power Maxwell} and other forms \cite{Hendi,Oliveira,Habib}. These solutions are mostly thought of as a model of a charged particle. However, since General Relativity is a nonlinear theory, one should study the stability of such models nonperturbatively. 

\fussy

Our aim here is to derive Robinson--Trautman solutions coupled to several forms of NE Lagrangians in order to investigate the influence of nonsphericity in this family on the results gained previously in highly symmetric situations. Since spherically symmetric solutions are a special sub case of Robinson--Trautman spacetime it is a natural family to consider for nonlinear stability investigation.
 
\section{Vacuum Robinson--Trautman solutions and field equations}
\label{RTmetricsec}

The general form of a vacuum Robinson--Trautman spacetime can be represented by the following line element \cite{RobinsonTrautman:1960,RobinsonTrautman:1962,Stephanietal:book,Griffiths:book}
\begin{equation}\label{RTmetric}
\d s^2_{RT}= -2H\,\d u^2-\,2\,\d u\,\d r + \frac{r^2}{{P}^2}\,(\d x^{2} + \d y^{2}),
\end{equation}
with ${2H = \Delta(\,\ln {P}) -2r(\,\ln {P})_{,u} -{2m/r} -(\Lambda/3) r^2}$,
\begin{equation}\label{Laplace}
\Delta\equiv {P}^2(\partial_{xx}+\partial_{yy}),
\end{equation}
and where $\Lambda$ is the cosmological constant. The metric depends on two functions, ${\,{P}(u,x,y)\,}$ and ${\,m(u)\,}$, which satisfy the nonlinear fourth-order PDE (so called Robinson--Trautman equation)
\begin{equation}
\Delta\Delta(\,\ln {P})+12\,m(\,\ln {P})_{,u}-4\,m_{,u}=0\,.
\label{RTequationgen}
\end{equation}
The function $m(u)$ might be set to a constant by a suitable coordinate transformation for vacuum solution. However, for a null radiation field source which is aligned with the principal null direction the solution represents generalization of Vaydia spacetime with time-dependent $m(u)$.



The spacetime is defined by geodesic, shearfree, twistfree and expanding null congruence generated by ${\boldk=\partial_r}$. The coordinate $r\in (0,\infty)$ is an affine parameter along this congruence, $u\in (u_{0},\infty)$~is a retarded time coordinate (the initial data for this class of spacetimes are specified on null hypersurface $u=u_{0}$), and $x,y$ are spatial coordinates spanning transversal 2-space which have Gaussian curvature (for ${r=1}$)
\begin{equation}
{K}(u,x,y)\equiv\Delta(\,\ln {P})\,.
\label{RTGausscurvature}
\end{equation}
If we select the transversal 2-spaces to be topological spheres (standard assumption in this class) then $x,y$ are real stereographic-type coordinates on a deformed spheres $r=const, u=const$. For general fixed values of $r$ and $u$, the Gaussian curvature is ${{K}/r^2}$ so that, as ${r\to\infty}$, they become locally flat.


\section{Robinson--Trautman and static spherically symmetric solutions}\label{Method}
\fussy
In this section we will develop a generating method to obtain a Robinson--Trautman solution coupled to NE based on a static spherically symmetric spacetime solution with NE. In order to have a straightforward generalization, we use the Kerr--Schild-type modification of the vacuum Robinson--Trautman metric. First, we find the Einstein equations for Robinson--Trautman spacetime coupled to NE and then compare them to SSS ones. Based on the similarities we formulate the theorem summarizing the generating method.
\sloppy

We consider the following action, describing nonlinear electrodynamics coupled to gravity, 
\begin{equation}\label{action}
	S=\int d^{4}x \sqrt{-g}[\mathcal{R}-2\Lambda+\mathcal{L}(F)]
\end{equation}
where $\mathcal{R}$ is the Ricci scalar for the metric $g_{\mu \nu}$ and Lagrangian of the nonlinear electromagnetic field $\mathcal{L}(F)$ is an arbitrary function of the electromagnetic field invariant $F=F_{\mu \nu}F^{\mu \nu}$  constructed from closed Maxwell 2-form $F_{\mu \nu}$. We use units in which $c=\hbar=8 \pi G=1$. By applying the variation with respect to the metric for the action (\ref{action}), we get Einstein equations
\begin{equation}\label{field equations}
	G^{\mu}{}_{\nu}=T^{\mu}{}_{\nu}-\Lambda \delta ^{\mu}{}_{\nu}.
\end{equation}
The energy momentum tensor generated by NE Lagrangian is given by 
\begin{equation}\label{energy-momentum-Maxwell}
	T^{\mu}{}_{\nu}=\frac{1}{2}\{\delta ^{\mu}{}_{\nu} \mathcal{L}-4\,(F_{\nu \lambda}F^{\mu \lambda})\mathcal{L}_F\}
\end{equation}
and the modified Maxwell (nonlinear electrodynamics) field equations are then given in the following form 
\begin{equation} \label{modified Maxwell}
	\partial _{\mu}(\sqrt{-g}\mathcal{L}_FF^{\mu \nu})=0
\end{equation} 
in which $\mathcal{L}_F=\frac{d\mathcal{L}(F)}{dF}$.

Our metric ansatz is
\begin{equation}\label{RT-Non}
	\d s^2 = \d s^2_{RT} -Q(u,r)\,\d u^2\ ,
\end{equation} 
where we modify the vacuum Robinson--Trautman metric, i.e. (\ref{RTmetric}), by adding function $Q(u,r)$ to the $g_{uu}$ component. This corresponds to general Kerr-Schild metric form given by background vacuum Robinson-Trautman geometry $g^{RT}_{\mu\nu}$ and null, shearfree, twistfree and geodetic vector $\boldk$
\begin{equation}
  g_{\mu\nu}=g^{RT}_{\mu\nu} - Q(u,r)k_{\mu}k_{\nu}
\end{equation}
Einstein tensor has now more nontrivial components compared to the original Robinson--Trautman metric (\ref{RTmetric}) 
\begin{eqnarray}
{G^{u}}_{u}&=&{G^{r}}_{r}=-\Lambda+\frac{Q_{,r}}{r}+\frac{Q}{r^2} \label{Guu}\\
{G^{x}}_{x}&=&{G^{y}}_{y}=-\Lambda+\frac{1}{2}{Q_{,rr}}+\frac{Q_{,r}}{r} \label{Gxx}\\
{G^{r}}_{u}&=&{}^{\mathrm{RT}}{G^{r}}_{u}-\frac{1}{r}[(\,\ln {P})_{,u} \{r{Q_{,r}}-2Q\}+{Q_{,u}}] \label{Gur}
\end{eqnarray}
where 
$${}^{\mathrm{RT}}{G^{r}}_{u}=-\frac{1}{2r^2}\{ \Delta\Delta(\,\ln {P})+12\,m(\,\ln {P})_{,u}-4\,m_{,u}\}$$

We assume the following specific Maxwell 2-form in the coordinates of (\ref{RT-Non})
 \begin{equation}\label{Twoform}
 	\mathbf{F}=F_{\mu\nu} \d x^{\mu} \d x^{\nu}=E(u,r) \d u \wedge \d r 
 \end{equation} 
 where the electromagnetic invariant $F=F_{\mu \nu}F^{\mu \nu}$ for the above Maxwell 2-form simplifies to $-2E^2$. Then from (\ref{modified Maxwell}) and the metric (\ref{RT-Non}) one can find dynamical equation for electromagnetic field
 \begin{equation} \label{NEQ-1}
 	\frac{r^2}{P(u,x,y)^2}\mathcal{L}_F F_{ur}=F_{0}(x,y),
 \end{equation} 
which can be solved by 
 \begin{equation} \label{NEQ-2}
 	\mathcal{L}_F F_{ur}= \frac{q(u)^2}{r^2},
 \end{equation}
 in which $P(u,x,y)^{2}F_{0}(x,y)=q(u)^2$ in order to satisfy the assumed form of $\mathbf{F}$ (\ref{Twoform}).
 The energy momentum tensor given by (\ref{energy-momentum-Maxwell}) can then be expressed in the diagonal form 
 \begin{equation}\label{energy-momentum-NE}
 	T^{\mu}{}_{\nu}=diag\left\{\frac{\mathcal{L}}{2}-F\mathcal{L}_F,\frac{\mathcal{L}}{2}-F\mathcal{L}_F,\frac{\mathcal{L}}{2},\frac{\mathcal{L}}{2}\right\}
 \end{equation}
for our form of Maxwell field (\ref{Twoform}) and with respect to the coordinates $(u,r,x,y)$ of metric (\ref{RT-Non}).
 
 Let us note that the assumption (\ref{Twoform}) necessarily leads to $F_{\mu \nu}F^{\mu \nu}\neq 0$ and $F^{*}_{\mu \nu}F^{\mu \nu}=0$ (where $*$ means a Hodge dual) as in the static spherically symmetric cases considered below. This also means that our field is non-null and purely electric. At the same time we do not need to consider generalization of Lagrangian that would additionally contain terms dependent on the second invariant.

Now, we will turn our attention to the general form of previously derived static spherically symmetric (SSS) solutions with NE. The metric has the following form encompassing all the models that we wish to generalize
\begin{eqnarray}\label{SSS}
\d s^2& =& -\left[1-{\textstyle{\frac{\Lambda}{3}}} r^2+f(r)\right]\,\d t^2+\,\frac{\d r^{2}}{\left[1-{\textstyle{\frac{\Lambda}{3}}} r^2+f(r)\right]} \nonumber \\
 &&+{r^2}\,(\d {\theta}^{2}+{{\sin}^{2} {\theta}} \, \d  {\phi}^{2}),
\end{eqnarray} 
Here, we prefer not to introduce the null coordinate in order to keep the correspondence with published spherically symmetric models. In this case one can of course as well use the Kerr-Schild form to introduce function $f$ into the background Minkowski/(anti-)de Sitter metric.

We obtain these nonzero components of the Einstein tensor for line element (\ref{SSS})
\begin{eqnarray}
{G^{t}}_{t}&=&{G^{r}}_{r}=-\Lambda+\frac{f_{,r}}{r}+\frac{f}{r^2} \label{SSSGtt}\\
{G^{\theta}}_{\theta}&=&{G^{\phi}}_{\phi}=-\Lambda +\frac{1}{2}{f_{,rr}}+\frac{f_{,r}}{r} \label{SSSGxx}
\end{eqnarray}

Maxwell 2-form has the following form resembling (\ref{Twoform}), 
\begin{equation}\label{Twoform-SSS}
	\mathbf{F}=\tilde{E}(r) \d t \wedge \d r 
\end{equation}
although the electromagnetic field is static now (like before, field invariant simplifies considerably $F=F_{\mu \nu}F^{\mu \nu}=-2\tilde{E}^{2}$). From the modified Maxwell equation (\ref{modified Maxwell}) and the metric (\ref{SSS}), we find 
\begin{equation} \label{NEQ-SSS}
	\mathcal{L}_F F_{tr}= \frac{q_{0}^2}{r^2},
\end{equation}
where $q_{0}$ is a constant. One can easily calculate the energy momentum tensor in this case and recovers the diagonal form (\ref{energy-momentum-NE}) but now with respect to the coordinates $(t,r,\theta,\phi)$ of metric (\ref{SSS}). 


We can see the structural similarity between set of equations (\ref{Guu},\ref{Gxx}) and (\ref{SSSGtt},\ref{SSSGxx}). This suggests trying to generalize SSS solutions with NE source to Robinson--Trautman geometry via this similarity. Note, that this similarity is not analyzed based on some coordinate transformation but purely from the perspective of coincidence between the form of differential equations. Of course this similarity has underlying reason in the fact that Robinson--Trautman metrics are generalization of spherically symmetric models but this is not important for the following observations. Namely, the solution $f(r)$ of equations (\ref{SSSGtt},\ref{SSSGxx}) can be evidently transformed into a particular solution $Q(u,r)$ of equations (\ref{Guu},\ref{Gxx}) by promoting integration constants in $f$ into functions of coordinate $u$. However, one has to ensure that the newly constructed function $Q(u,r)$ satisfies additional equation ${G^{r}}_{u}=0$ coming from combining corresponding components of Einstein tensor (\ref{Gur}) and energy momentum tensor (\ref{energy-momentum-NE}) into field equations (\ref{field equations}). This equation is the basic constraint on generalizing SSS solutions with NE into Robinson--Trautman case.

We summarize the method in the following theorem:
\begin{theorem}\label{theorem}
  The SSS solution of Einstein equations coupled to arbitrary NE model which is given by line element in the form of (\ref{SSS}) and Maxwell field (\ref{Twoform-SSS}) with specific function $f(r)$ can be generalized into Robinson--Trautman solution coupled to the same NE model with metric (\ref{RT-Non}) and Maxwell field (\ref{Twoform}) where the function $Q(u,r)$ is obtained from $f(r)$ by promoting the integration constants (appearing in $f$) into functions of $u$ provided the additional constraint equation
  \begin{equation}\label{constraint}
    (\,\ln {P})_{,u} \{r{Q_{,r}}-2Q\}+{Q_{,u}}=0
  \end{equation}
  is satisfied for such $Q(u,r)$.
\end{theorem}

\section{Explicit examples}
Now, we will present several important forms of NE Lagrangians and the associated solutions for both electromagnetic field and Robinson--Trautman metric (giving the form of metric function $Q(u,r)$). For each model we first solved the spherically symmetric equations (\ref{SSSGtt}) and (\ref{SSSGxx}) to obtain the most general function $f(r)$, then used the Theorem \ref{theorem} to generate $Q(u,r)$ with which we try to solve the constraint (\ref{constraint}). We do not present the original function $f(r)$ since it can be easily read of from $Q(u,r)$ by putting all functions of $u$ to constants. Finally, we specify in which references one can find the spherically symmetric case.

\begin{figure}[t]
\centering 
\includegraphics[trim = 25mm 140mm 80mm 20mm,scale=0.7]{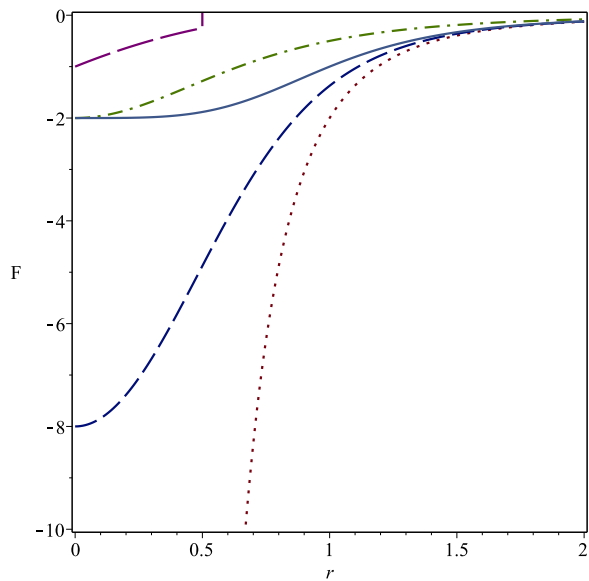}
\caption{The plot of electromagnetic field invariant $F$ for Maxwell (dotted), Born--Infeld (solid line), Logarithmic(dashed), New Lagrangian 1 (dash-dotted), New Lagrangian 2 (long-dashed) models. The free parameter was fixed to following value: $\beta=1$. We have also fixed the function $q$ to a constant $q(u)=1$ which can be considered as $u=const$ slice of the complete three-dimensional plots given in Appendix. The graph corresponding to New Lagrangian 2 is limited only to the coordinate range where the model satisfies energy conditions. Only in the Maxwell case the invariant $F$ diverges at origin.}
    \label{Fs}
\end{figure}

We plotted the profile of Lagrangian (Figure \ref{Lagrangians}) for all the models considered below and electromagnetic field invariant $F$ (Figure \ref{Fs}) in case of static spherically symmetric solutions (alternatively it can be viewed as a profile of invariant on $u=const$ hypersurface). Evolution of the invariant for Robinson--Trautman solutions is presented in three-dimensional plots in Appendix.

\subsection{Maxwell}
As a starting point we briefly show the linear Maxwell case which generalizes the standard Reissner-Nordstr\"{o}m solution 
\begin{equation} 
\mathcal{L}(F)=-F
\end{equation}
Applying the method described in section \ref{Method} and summarized in Theorem \ref{theorem} we obtain this solution on Robinson--Trautman background
\begin{eqnarray}
Q(u,r)&=&\frac{q(u)^4}{r^2}+\frac{C_{1}(u)}{r}
\end{eqnarray}
If we put this form into the field equation constraint (\ref{constraint}) we find 
\begin{eqnarray}\label{restrictions-Max}
\Delta\Delta(\,\ln {P})+12\,M(\,\ln {P})_{,u}-4\,M_{,u}&=&0 \nonumber \\
 q\, P_{,u}-q_{,u}\,P &=&0 
\end{eqnarray}
where
\begin{equation}\label{Mfunction}
 M=m(u)-\frac{C_{1}(u)}{2}.
\end{equation}
This result is a special case of algebraic type II Einstein--Maxwell spacetimes analyzed in \cite{Stephanietal:book} where a theorem (Theorem 28.3 in the reference) for generating these solutions from the vacuum Robinson--Trautman spacetime is presented.

\subsection{Born-Infeld}
As was mentioned before, NE started with this model of dynamics and it has been studied widely in three, four and higher dimensions, also their physical properties were analyzed in detail. The famous Born--Infeld Lagrangian has the following form \cite{BornInfeld}
\begin{equation} \label{Born-Infeld}
\mathcal{L}_{BI}=4\beta^2\left(1-\sqrt{1+\frac{F}{2\beta^2}}\right) ,
\end{equation}
 $\beta $ is a critical field length. When $\beta \rightarrow \infty $ this Lagrangian goes to Maxwell form which is as well recovered in the weak field regime.
 
\begin{figure}[t]
\centering 
\includegraphics[trim = 25mm 140mm 80mm 20mm,scale=0.7]{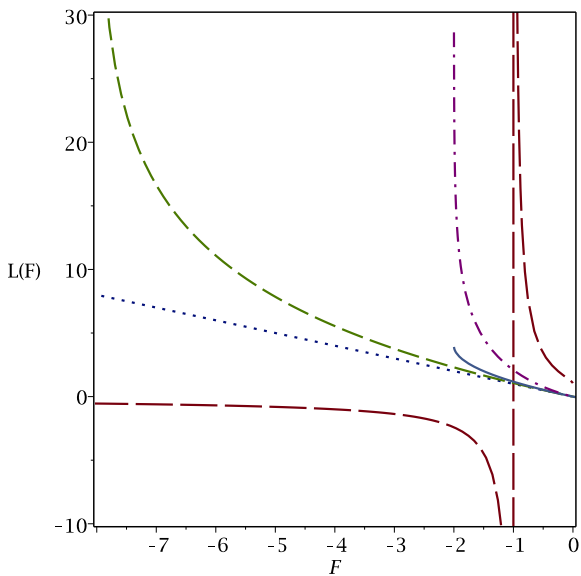}
\caption{The plot of all considered Lagrangians for NE: Maxwell (dotted), Born--Infeld (solid line - it is defined only for restricted values of $F$), Logarithmic model (dashed), New Lagrangian 1 (dash-dotted), New Lagrangian 2 (long-dashed). The free parameter was fixed to following value: $\beta=1$. Evidently, all the nonlinear Lagrangians have steeper growth with increasing absolute value of $F$ than Maxwell model and as well either diverge for certain finite value of $F$ or fail to be defined (Born--Infeld). Except for the New Lagrangian 2 model the value at $F=0$ is the same for all models.}
    \label{Lagrangians}
\end{figure}

Applying the method described in Section \ref{Method} we obtain this solution on Robinson--Trautman background
\begin{equation}\label{F-Born-Infeld} 
F=-\frac{2q(u)^4}{r^4+{q(u)^4}/{\beta^2}}.
\end{equation} 
As it was expected the electric field remains regular at $r=0$. Lagrangian (\ref{Born-Infeld}) is only defined for field invariant $F\geq -2\beta^{2}$ (see as well Figure \ref{Lagrangians}) which corresponds to $r\geq 0$ according to (\ref{F-Born-Infeld}) so it covers the whole range of coordinates. The metric function corresponding to NE source becomes
\begin{eqnarray}
Q(u,r)=\frac{C_{1}(u)}{r} +\frac{2\beta^2}{3}r^2-\frac{2\beta^2}{r}\int{\sqrt{r^4+{q(u)^4}/{\beta^2}}}\,dr\nonumber\\
\end{eqnarray}
If we put these forms in field equations we will find the same restrictions as for Maxwell theory (\ref{restrictions-Max}). However, in this case the constraint (\ref{constraint}) enforces $C_{1}(u)=0$ and so $M=m$.

Such solution is a straightforward generalization of the SSS solutions given for example in \cite{Fernando,Oliveira}. This shows that there is indeed a solution in Robinson--Trautman class that settles down smoothly to known spherically symmetric one when function $P$ attains its special form describing the geometry on a sphere (this is elaborated in Section \ref{Asymptotics}). Then, from (\ref{restrictions-Max}) one immediately concludes that $q=const$. 

\subsection{Logarithmic form of nonlinear electrodynamics theory}
The logarithmic form of Lagrangian can be given by
 \begin{equation} \label{logarithmic}
\mathcal{L}_{LNE}=-8\beta^2 \ln {\left({1+\frac{F}{8\beta^2}}\right) }.
\end{equation}
In \cite{Soleng} it was used to present a model of point particle without divergence in electromagnetic field. However, even with vanishing "Schwarzschild mass" and finite electromagnetic field the curvature singularity at the origin is present. For special value of $\beta$ the horizon radius shrinks to zero and so called black point is created (see, e.g., \cite{black-point} for their previous occurrences).

Applying the method described in section \ref{Method} we obtain
\begin{equation}\label{F-logarithmic}
	F=-\frac{8\beta^2}{q(u)^4}\,{\left(\beta r^2-\sqrt{\beta^2r^4+q(u)^4}\right)}^2\ .
\end{equation} 
The presence of curvature singularity can be understood from Lagrangian (\ref{logarithmic}) since it has singularity for field invariant value $F= -8\beta^{2}$ which corresponds to $r=0$ according to (\ref{F-logarithmic}) and so this Lagrangian provides singular source for Einstein equations. The Robinson--Trautman geometry is specified by the following metric function
\begin{eqnarray}
Q(u,r)=\frac{C_{1}(u)}{r} +\frac{20}{9}\beta^2r^2-4 \beta \sqrt{r^4\beta^2+q(u)^4} \nonumber\\
-\frac{4}{3} \beta^2 r^{2}\ln \left( {\frac{-2r^4 \beta^2+2 \beta r^2 \sqrt{r^4\beta^2+q(u)^4}}{q(u)^4}}\right) \nonumber\\
+\frac{16 \beta^3}{r} \int{\frac{r^4}{\sqrt{r^4\beta^2+q(u)^4}}}\,dr
\end{eqnarray}
If we put these $Q$ and $F$ into the constraint (\ref{constraint}) we recover conditions (\ref{restrictions-Max}) and (\ref{Mfunction}).

This solution is again a straightforward generalization of the SSS solutions given in \cite{Soleng}. The results concerning spherically symmetric limit are the same as in the previous analysis of Born--Infeld example.  

\subsection{New Lagrangian 1}
Now we will consider another form of dynamics for NE given by the following Lagrangian \cite{Habib3}
 \begin{equation} \label{New}
\mathcal{L}_{New}=-\frac{2}{\alpha^4} \ln{\left(1-{\alpha^2}\sqrt{-{F}}\right)}-\frac{2\sqrt{-{F}}}{\alpha^2},
\end{equation}
which has correct Maxwell limit in both weak field regi\-me and for $\alpha\to 0$.
The generating method from section \ref{Method} yields the solution 
\begin{equation} 
F=-\frac{2q(u)^4}{\left(r^2+q(u)^2\beta^2\right)^2}
\end{equation} 
in which $\beta^2=\sqrt{2} \alpha^2 $. Again, the field is regular however the Lagrangian has singularity for $F=-\alpha^{-4}$ (see Figure \ref{Lagrangians}) which is attained at $r=0$. Note that in this case one needs to select negative root when solving for $E(r,u)$ in order to satisfy the field equations. Metric function becomes
\begin{eqnarray}
Q(u,r)&=&\frac{C_{1}(u)}{r}-\frac{2}{3}\frac{q(u)^2}{\beta^2}-\frac{4}{3}\frac{q(u)^3}{r\beta}\arctan{\left(\frac{r}{q(u)\beta}\right)} \nonumber\\
&&-\frac{2r^2}{3\beta^4}\ln{\left(\frac{r^2}{r^2+q(u)^2\beta^2}\right)}
\end{eqnarray}
If we put these forms in field equations we will find the already known set of restrictions (\ref{restrictions-Max}) and (\ref{Mfunction}). This solution generalizes the SSS solutions given in \cite{Habib3} where this type of Lagrangian is used for the first time.

\subsection{New Lagrangian 2}
Finally, we consider the following NE Lagrangian \cite{Habib2}
\begin{equation} \label{New2}
\mathcal{L}_{New2}=\frac{1}{\beta-\sqrt{-{F}}},
\end{equation}
which does not have a Maxwell limit in the weak field regime. However, models containing square root (or arbitrary powers) and generally devoid of Maxwell limit were extensively discussed before \cite{square root,power Maxwell}. And these examples can serve as an approximation of true dynamics in strong field regime. Additional motivation to consider a model without Maxwell limit is to test whether the generating method used here is not limited (via the constraint (\ref{constraint})) only to those NE models with correct Maxwell limit (which are all those considered so far).

The generating method from section \ref{Method} yields the solution
\begin{equation} \label{F-New2}
F=-\frac{\left(\beta\,\,q(u)-r\right)^2}{q(u)^2}
\end{equation} 
and we need $q(u)>0,\ \beta q(u)-2r>0$ to fulfill the weak energy condition so this solution cannot cover the whole coordinate range but can serve as an inner solution for small $r$. Evidently, without introducing the parameter $\beta$ we cannot satisfy the energy conditions at all (for more discussion and interpretation see \cite{Habib2}).

The geometry is in this case defined by
\begin{eqnarray}
Q(u,r)&=&\frac{C_{1}(u)}{r}-\frac{\beta q(u)^2}{2}+\frac{q(u)}{2}r\ .
\end{eqnarray}
If we put these forms in field equations we will find the already known set of restrictions (\ref{restrictions-Max}) and (\ref{Mfunction}). This solution is a straightforward generalization of the SSS solutions given for example in \cite{Habib2}.

The Lagrangian (\ref{New2}) has a singularity at $\sqrt{-F}=\beta$ (see as well the Figure \ref{Lagrangians}) which corresponds to $r=0$ according to (\ref{F-New2}). This means that although the field invariant $F$ is regular the curvature singularity is created via singular behavior of the Lagrangian.

\section{Asymptotic behavior}\label{Asymptotics}
The metric (\ref{RTmetric}) admits coordinate freedom already noted by Robinson and Trautman
\begin{equation}\label{transform}
u'=U(u), \, \, r'=\frac{r}{U_{,u}}, \, \, m'=\frac{m}{U_{,u}^3},\,\,P'=\frac{P}{U_{,u}},
\end{equation}
which can be used to set mass $M'$ to a positive constant by proper function $U(u)$. In the case of nonlinear electrodynamics, when considering metric (\ref{RT-Non}) and modified Maxwell equations (\ref{modified Maxwell}), one has to supplement (\ref{transform}) (now with $M=m-\frac{C_{1}}{2}$ instead of $m$) by transforming $q(u)$ as well
\begin{equation}
q'=\frac{q}{U_{,u}},
\end{equation}
Note that one can not set $q(u)$ and $M(u)$ to constant simultaneously since one has only single function in hand. This means that $q(u)$, although looking like physical charge based on (\ref{NEQ-2}), shares the same interpretation problems as $m(u)$. Only now these difficulties are combined together.

Now we are ready to investigate asymptotic behavior for our solutions, separately for the retarded time $u\to\infty$ and $r\to\infty$:{}

\subsection{Asymptotics $u\to\infty$}
One can use (\ref{transform}) to put $M'$ to constant in (\ref{Mfunction}) for all models exactly the form considered in Chru\'{s}ciel and Singleton \cite{Chru1,Chru2,ChruSin} analysis of asymptotic behavior to recover the spherically symmetric final state and also exponentially fast decay of dependence of function $P'$ on new coordinate $u'$. Due to the last equation in (\ref{restrictions-Max}), the asymptotic behavior of function $q'$ is the same as $P'$. Namely, it tends to a constant which is completely consistent with the final state approaching the corresponding spherically symmetric solution.

\subsection{Asymptotics $r\to\infty$}
In this limit the electromagnetic field vanishes for all cases and for Born--Infeld, Logarithmic and New Lagrangian 1 models the asymptotic behavior is identical to the Maxwellian case. Analyzing behavior of function $Q(u,r)$ (after moving $C_{1}$ term into redefinition of "mass" $m\to M$) we immediately see that in all cases the metric is locally asymptotically flat (or (anti--)de Sitter) as for the vacuum Robinson--Trautman solution. We have neglected the New Lagrangian 2 model due to its restricted coordinate range.

\section{Horizons}
Since in all cases the Robinson--Trautman spacetimes with arbitrary form of electrodynamics still posses curvature singularity at $r=0$, as one can confirm by computing Kretschmann scalar (singularity is milder for NE models with regular EM field at the origin), we would like to know if it is covered by a horizon. Due to the dynamical nature of our spacetimes we will look for quasilocal horizon. So we need to find a marginally trapped surfaces and one can select any of the most popular horizon definitions --- apparent \cite{hawking-ellis}, trapping \cite{hayward} or dynamical horizon \cite{krishnan}. We will be looking for a horizon hypersurface given by the equation $r=\mathcal{N}(u,x,y)$ with the $u=u_{0}=const$ slices
\begin{equation}
	r=\mathcal{N}(u_{0},x,y)=N(x,y)
\end{equation}
being marginally trapped surfaces. We shall investigate the expansions of both null normals to this surface
\begin{eqnarray}\label{null-vectors}
	\mathbf{{k}}&=&\partial_{r}\\
	\mathbf{{l}}&=&\partial_{u}+\left[\frac{P^{2}}{r^{2}}(N_{,x}^{2}+N_{,y}^{2})-H-\frac{Q}{2}\right]\partial_{r}+\nonumber\\
	&&\frac{P^{2}}{r^{2}}(N_{,x}\partial_{x}+N_{,y}\partial_{y})\nonumber
\end{eqnarray}
that are normalized using $g(\mathbf{{l}},\mathbf{{k}})=-1$. The congruence generated by $\mathbf{{k}}$ is the one defining Robinson--Trautman family and thus apart from being shearfree and twistfree it has positive expansion everywhere $\Theta_{\mathbf{{k}}}>0$. So by demanding the other expansion to vanish $\Theta_{\mathbf{{l}}}=0$ we are looking for the past horizon according to the definition by Hayward \cite{hayward}. One can express this expansion in the following form
\begin{eqnarray}
	\Theta_{\mathbf{{l}}}&=&\frac{-1}{r}\left[K-\frac{2m}{r}-\frac{\Lambda}{3}r^{2}+Q\right.\\
	&&\qquad\quad\left.-\frac{r\Delta N-P^{2}(N_{,x}^{2}+N_{,y}^{2})}{r^{2}}\right]\nonumber
\end{eqnarray}
which leads upon evaluation on the horizon surface to equation
\begin{equation}\label{horizon-eq1}
	K-\frac{2m}{N}-\frac{\Lambda}{3}N^{2}-\Delta\ln N +Q(u_{0},N)=0
\end{equation}
where only the last term represents generalization of the horizon equation derived in \cite{PodSvi:2009}. We can check using expansion at origin and infinity that $Q(u,r)$ (after moving the $C_{1}$ term into redefinition of "mass") is regular everywhere for all NE models considered above while it naturally diverges at origin for Maxwell theory. For the last NE model this regularity is in fact caused by the restricted range of coordinate for $r$.

\sloppy

First, let us use the following redefinition of the function describing horizon $N(x,y)=2mc\e^{-\Phi(x,y)}$ given in \cite{PodSvi:2009} to obtain an equation (\ref{horizon-eq1}) in better form
\begin{equation}\label{mastereq}
	\Delta\Phi=\frac{1}{c}\,\e^{\Phi}+\frac{4}{3}\Lambda m^2 c^2\,\e^{-2\Phi}-K-Q(u_{0},2mc\e^{-\Phi})   \,,  
\end{equation}
where we assume $\Phi>0$.{}

\fussy

In the case of NE models the regularity of $Q$ at origin and infinity (or maximum allowed value of $r$) means that it has finite supremum and infimum 
\begin{equation}
Q_{\sup}=\sup_{r\in (0,\infty)}Q(u_{0},r)\, ;\ Q_{\inf}=\inf_{r\in (0,\infty)}Q(u_{0},r)
\end{equation}
These, together with minimum $K_{\min}$ and maximum $K_{\max}$ of Gaussian curvature of compact surface spanned by $x$ and $y$, can be used to straightforwardly generalize the results given in \cite{PodSvi:2009} which use theorem (Theorem 1 therein) relying on the existence of sub- and super-solutions \cite{footnote} $\Phi_{\pm}$ for (\ref{mastereq}) satisfying $0<\Phi_{-}\leq \Phi_{+}$. In our case the constant sub- and super-solutions can be given depending on the value of cosmological constant:
\begin{itemize}
	\item[$\bullet$] $\Lambda \leq 0$
	\begin{eqnarray}
		\Phi_{-} &=& \ln \left( \,c\, [K_{\min}+Q_{\inf}]\, \right) \, \nonumber\\
		\Phi_{+} &=& \ln \Big( \,c\, [K_{\max}+Q_{\sup}] - \frac{4}{3}\Lambda m^2 c^3\, \Big) \,   \label{subsuper}
	\end{eqnarray}
	provided $c[K_{\min}+Q_{\inf}]>1$,
	\vspace{8pt}
	\item[$\bullet$] $\Lambda > 0$
	\begin{eqnarray}
		\Phi_{-} &=& \ln \Big( \,c\, [K_{min}+Q_{\inf}] - \frac{4}{3}\Lambda m^2 c^3\, \Big) \, \nonumber\\
		\Phi_{+} &=& \ln \left( \,c\, [K_{max}+Q_{\sup}]\, \right) \,   \label{subsuper+}
	\end{eqnarray}
	if ${\,c\, [K_{min}+Q_{\inf}] - \frac{4}{3}\Lambda m^2 c^3 >1 }$. By using the optimal choice of constant $c$ this constraint reduces to condition on "physical" quantities
	\begin{equation}
		9\Lambda m^{2} < (K_{min}+Q_{\inf})^{3}.
	\end{equation}
\end{itemize}

\sloppy

So the restrictions on existence of horizon are stronger for positive cosmological constant which is natural since as a special case for $Q=0$ we have asymptotically (there we have $K_{min}=1$) Schwarzschild--de Sitter solution which can have naked singularity (see \cite{PodSvi:2009} for extended discussion).

\fussy

For the Maxwell theory one has to be more careful when constructing sub- and super-solutions. The sub-solutions for both cases of $\Lambda$ can be used here straightforwardly by setting $Q_{\inf}=0$. For super-solution one uses the explicit form $Q=q(u_{0})^{4}/r^{2}$ to derive quadratic equation for $z=\e^{\Phi}$
\begin{equation}\label{z-equation}
	\frac{1}{c}\,z+\frac{4}{3}\Lambda m^2 c^2-K_{\max}-\frac{q(u_{0})^{4}}{4m^{2}c^{2}}z^{2}=0
\end{equation}
based on the right-hand side of (\ref{mastereq}) after using upper bound for the second and third term. Upon finding positive solution of (\ref{z-equation}) for an optimal choice of free constant $c$ one can give the following super-solutions (notice slightly different division of cases according to $\Lambda$):
\begin{itemize}
	\item[$\bullet$] $\Lambda < 0$
	\begin{equation}
		\Phi_{+} = \ln \left( \frac{2m^{2}c_{1}}{q(u_{0})^{4}}\right) \, \nonumber\\
	\end{equation}
	where $c_{1}=-\frac{\sqrt{3\Lambda(K_{\max}q(u_{0})^{4}-m^{2})}}{2\Lambda q(u_{0})^{2}m}$, so the necessary condition is 
	\begin{equation}\label{condition-Maxwell}
	m^{2}>K_{\max}q(u_{0})^{4}
	\end{equation}
	\vspace{8pt}
	\item[$\bullet$]$\Lambda \geq 0$
	In this case one can neglect the cosmological constant term since it has the preferable sign anyway and directly obtain
	\begin{equation}
		\Phi_{+} = \ln \left(\frac{2mc}{q(u_{0})^{4}}(m+\sqrt{m^{2}-K_{\max}q(u_{0})^{4}}) \right) \, \nonumber\\
	\end{equation}
	and the condition is same as in the previous case (\ref{condition-Maxwell}).
\end{itemize}

\sloppy

Asymptotically $K_{\max}\to 1$ since the solution tends to spherically symmetric case and $q(u)$ approaches constant $q_{0}$. If one would select traditional notation $q_{0}^{4}=Q^{2}$ (here and only here $Q$ denotes charge of Reisner--Nordstr\"{o}m solution) one recovers natural condition $m^{2}>Q^{2}$.

\fussy

\section{Algebraic type of the solution}
Now, we would like to see if the geometry of our spacetime is sufficiently general. Since vacuum Robinson--Trautman spacetime is generally of algebraic type II we would like our solution to be at least of the same type and not more special. Our preferred tetrad for determining the Weyl scalars of our solution is given by different null vectors compared to (\ref{null-vectors})
\begin{eqnarray}
	\mathbf{\tilde{k}}&=&\partial_{r}\nonumber\\
	\mathbf{\tilde{l}}&=&\partial_{u}-(H+Q/2)\partial_{r}\\
	\mathbf{\tilde{m}}&=&\frac{P}{\sqrt{2}r}(\partial_{x}+i\partial{y})\nonumber
\end{eqnarray}
where $i$ is a complex unit. The Weyl spinor computed from this tetrad has only the following nonzero components
\begin{eqnarray}\label{Weyl}
	\Psi_{2}&=&\frac{1}{12}\left[Q_{,rr}-\frac{2}{r}Q_{,r}+\frac{2}{r^{2}}Q\right]-\frac{m}{r^3}\nonumber\\
  \Psi_{3}&=&-\frac{\sqrt{2}P}{4r^{2}}(K_{,x}-iK_{,y})\\
  \Psi_{4}&=&\frac{1}{4r^{2}}\left[\{P^2(K_{,x}-iK_{,y})\}_{,x}-i\{P^2(K_{,x}-iK_{,y})\}_{,y}\right]\nonumber
\end{eqnarray}
Now, we can easily determine the type irrespective of possible non-optimal choice of tetrad by using the review of explicit methods for determining the algebraic type in \cite{Zakhary} that are based on \cite{Penrose}. Namely, when we use invariants
$$I=\Psi_{0}\Psi_{4}-4\Psi_{1}\Psi_{3}+3\Psi_{2}^{2},\ J={\rm det}\left(\begin{array}{ccc}
\Psi_{4} & \Psi_{3} & \Psi_{2}\\
\Psi_{3} & \Psi_{2} & \Psi_{1}\\
\Psi_{2} & \Psi_{1} & \Psi_{0}
\end{array}\right)$$
we can immediately confirm that $I^{3}=27J^{2}$ is satisfied so that we are dealing with type II or more special. At the same time generally $IJ\neq 0$ so it cannot be just type III. Moreover, the spinor covariant $R_{ABCDEF}$ has nonzero components
\begin{eqnarray}
	R_{011111}&=&\frac{1}{2}\Psi_{2}(2\Psi_{3}^{2}-3\Psi_{2}\Psi_{4})\\
	R_{111111}&=&\Psi_{3}(2\Psi_{3}^{2}-3\Psi_{2}\Psi_{4})
\end{eqnarray}
which means that generally the spacetime cannot be of type D. So indeed our NE solution is of the most general type possible for the Robinson--Trautman vacuum class. Which does not mean that there cannot be a NE solution of type I when one considers completely general Maxwell tensor ${\mathbf F}$. Moreover, inspecting the components of Weyl spinor (\ref{Weyl}) one concludes that in the special case of $K(x,y)=const > 0$ (constant positive Gaussian curvature of compact two-space spanned by $x,y$) the algebraic type becomes D consistent with spherical symmetry. Finally, since $\Psi_{3}=0$ implies $\Psi_{4}=0$ we cannot have all components of spinor covariant $Q_{ABCD}$ (see \cite{Penrose} and \cite{Zakhary}) vanishing while having nonvanishing Weyl spinor. This means that our family of solutions does not contain type N geometries.

\section{Conclusion and final remarks}

\sloppy

We have derived Robinson--Trautman solutions with source given by nonlinear electrodynamics for several specific models of NE Lagrangian (both with Maxwell limit and without). The solutions were derived based on known spherically symmetric ones by a method described in Section \ref{Method}. The Maxwell case was included for comparison as well. In all cases of NE the singularity of electromagnetic field is resolved as in the static spherically symmetric cases. However, it was not possible to satisfy additional constraint for having Robinson--Trautman solution with Hoffmann--Born--Infeld model or with NE model which provides source of Bardeen black hole. Both these models can be used to construct spherically symmetric solutions without curvature singularity. The impossibility to generalize these models in the absence of this symmetry suggests that this kind of resolution of curvature singularity might not be stable under nonlinear perturbations (at least within the Robinson--Trautman class). However, the Robinson--Trautman class does not contain rotating black holes (due to twist-free condition) and therefore our results do not need to be universally valid. Unfortunately, the twisting class of solutions does not permit analysis on the level presented here (there are no asymptotic behavior studies in the dynamical regime).

\fussy

\sloppy

Since in all models the curvature singularity is present we analyzed the existence of horizons using the quasilocal concepts. All solutions are generally of algebraic type II and asymptotically in retarded time approach their spherically symmetric versions. All models with unrestricted coordinate range also remain locally asymptotically flat (or (anti--)de Sitter) as their vacuum counterparts.

\fussy

The interpretation of "charge" $q(u)$ suffers from the same difficulties as that of "mass" in vacuum Robinson--Trautman solutions. The asymptotic behavior of $q$ in all our models is identical to that of function $P$ which describes geometry of two spaces of constant $u$ and $r$, namely in preferred coordinate $u$ it settles exponentially fast to a constant.

\section*{Appendix}
  Here we plot graphs of electromagnetic field invariant $F$ for all the considered models. We have put $C_{1}(u)=1, \beta=1, \alpha=2^{-1/4}$ and we have selected exponential decay behavior for function $q(u)=1+\e^{-u}$ which corresponds to the asymptotic behavior derived in Section \ref{Asymptotics}. The plots show that, except for the Maxwell case, $F$ is finite at $r=0$ for all models. The difference is only in the rate of approach to this finite value. All the models give evidently asymptotically ($r\to \infty$) vanishing $F$ and also fast exponential decay to SSS form of $F$ is clear.
  
  \begin{figure}[h]
\centering 
\includegraphics[trim = 25mm 140mm 80mm 20mm,scale=0.7]{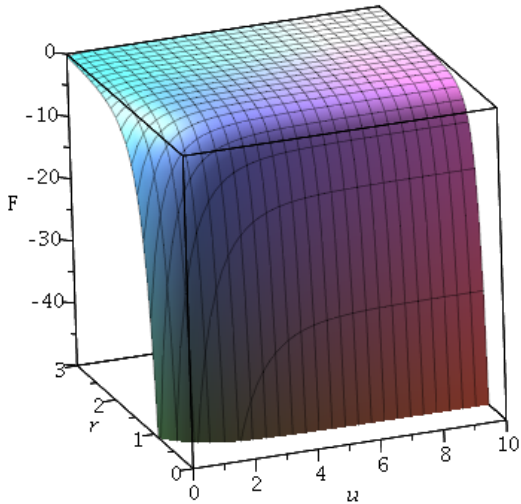}
\caption{The plot for Maxwell model}
    \label{F1}
\end{figure}

\begin{figure}[h]
\centering 
\includegraphics[trim = 25mm 140mm 80mm 20mm,scale=0.7]{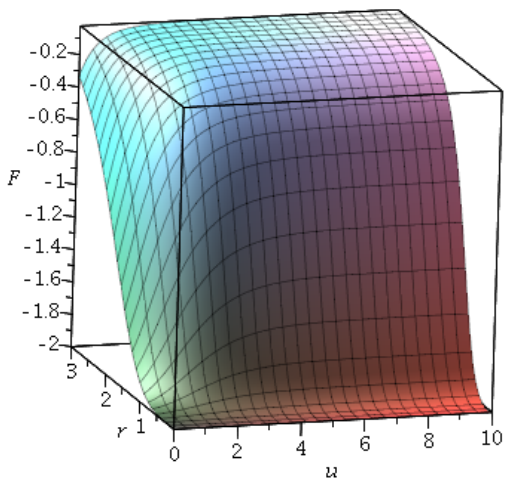}
\caption{The plot for Born--Infeld model}
    \label{F2}
\end{figure}

\begin{figure}[h]
\centering 
\includegraphics[trim = 25mm 140mm 80mm 20mm,scale=0.7]{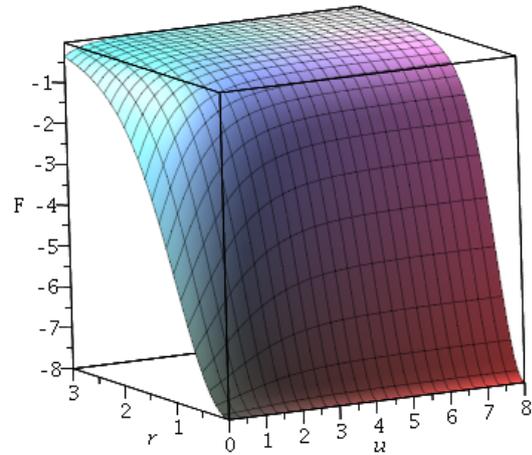}
\caption{The plot for Logarithmic model}
    \label{F3}
\end{figure}

\begin{figure}[h]
\centering 
\includegraphics[trim = 25mm 140mm 80mm 20mm,scale=0.7]{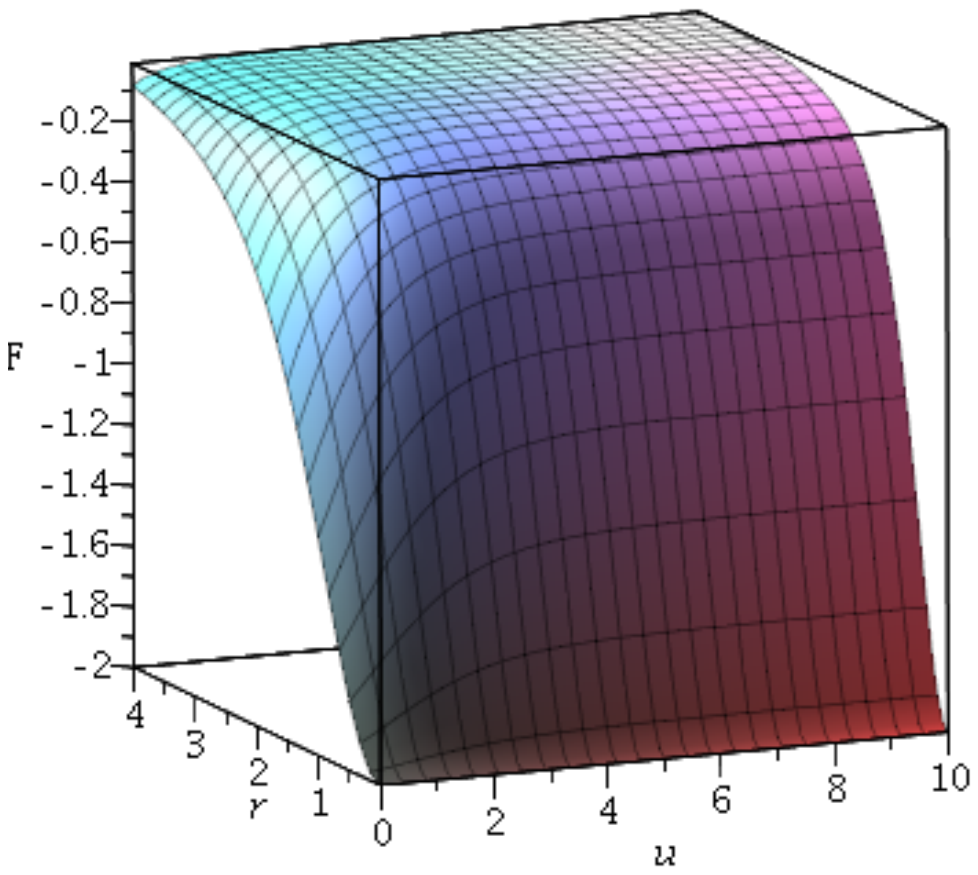}
\caption{The plot for New Lagrangian 1 model}
    \label{F4}
\end{figure}

\begin{figure}[h]
\centering 
\includegraphics[trim = 25mm 140mm 80mm 20mm,scale=0.7]{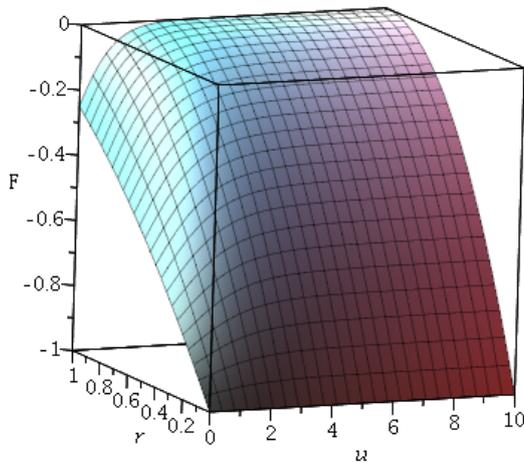}
\caption{The plot for New Lagrangian 2 model}
    \label{F5}
\end{figure}

\begin{acknowledgements}
We are grateful to prof. Ji\v{r}\'{\i} Bi\v{c}\'{a}k for discussion and valuable comments. This work was supported by grant GA\v{C}R 14-37086G. 
\end{acknowledgements}

\end{document}